\documentstyle[epsfig]{elsart}
\begin{document}
\begin{frontmatter}
\title{Limits on the low energy antinucleon-nucleus annihilations 
from the Heisenberg principle.} 
\author[Brescia]{A.~Bianconi},
\author[Brescia]{G.~Bonomi},
\author[Brescia]{M.P.~Bussa},
\author[Brescia]{E.~Lodi Rizzini},
\author[Brescia]{L.~Venturelli},
\author[Brescia]{A.~Zenoni},
\address[Brescia]{Dip. di Chimica e Fisica per 
l'Ingegneria e per i Materiali, 
Universit\`{a} di Brescia and
INFN, Sez. di Pavia, Italy} 

\begin{abstract}
We show that the quantum uncertainty principle puts  
some limits on the effectiveness of the antinucleon-nucleus 
annihilation at very low energies. This is caused by the fact 
that the realization a very effective short-distance reaction 
process implies information on the relative distance of the 
reacting particles. Some quantitative predictions are possible 
on this ground, including the approximate $A$-independence of 
$\bar{N}$-nucleus annihilation rates.   
\end{abstract} 
\end{frontmatter} 

\section{Introduction} 

Recently several experimental 
data\cite{obe1,obe2,wid,pro2,ne991,noi3} 
have shown that at projectile momenta below 200 MeV/c 
the behavior of antinucleon-nucleus annihilations is 
quite different from what could be naively expected. 

For $k$ (incident momentum of the antinucleon in the 
laboratory) below 70 MeV/c there are no 
evident signs of an increase of the $\bar{p}-$nucleus 
total annihilation cross section at increasing mass number $A$ 
of the target\cite{obe2,pro2,ne991}. At 30-50 MeV/c the 
$\bar{p}p$ total annihilation rate is larger than 
the corresponding rate for $\bar{p}$D and $\bar{p}^4$He. 
The width and shift of the ground level  
of antiprotonic atom of Hydrogen are larger than the 
corresponding observables in antiprotonic Deuterium\cite{wid}. 
For the 
$\bar{p}p$ scattering length $\alpha$ $\equiv$ $\alpha_R+i\alpha_I$ 
we have\cite{pro2,noi3,bat90} 
$\alpha_R$ $\approx$ $-\alpha_I$ $\approx$ 0.7$\div$0.8 fm, 
and the $\rho$-parameter (i.e. 
the ratio between the real and the imaginary part of 
the forward scattering amplitude) is $\sim$ $-1$ at 
zero energy. 
These values mean\cite{noi3} 
that at small momenta the elastic interaction is  
$repulsive$ (i.e. negative phase shifts: the outgoing 
scattered wave is in advance with respect to the free motion 
wave) and as much important as the annihilation. 

Elastic and annihilation data for $\bar{p}p$ at laboratory 
momenta $k$ below 600 MeV/c, scattering length data and 
$\rho-$parameter data from 0 to 600 MeV/c can be well 
fitted by energy-independent 
optical potentials\cite{noi3}. They present some $very$ 
curious features: (i) an increase of the strength of the 
imaginary part leads to a $decrease$ of the consequent reaction rate,  
and an increase in the radius of the imaginary part 
does not lead to a consistent increase of the reaction rate;  
(ii) a repulsive elastic amplitude is produced despite 
the real part of the potential is attractive; (iii) the 
annihilation rate is much more sensitive to the diffuseness 
parameter than to the strength or the radius. 
All this happens for $k$ $<$ 200 MeV/c. We could 
suspect that strange phenomena start from $k$ $\approx$ 200 MeV/c,  
although they become experimentally evident at smaller momenta. 

The synthesis of the previous facts can be: 
stronger and attractive in principle $\rightarrow$ 
weaker and repulsive in effect. 
We and other authors\cite{wyc,pro1,noi1,noi2} 
have presented explainations of these phenomena that 
for simplicity we regroup under the name ``inversion''. 
In particular, in \cite{wyc} it has been shown that, within a 
multiple scattering framework, double interaction terms 
interfere destructively with single interaction terms 
in $\bar{p}-$D interaction. In \cite{pro1} it has 
been shown that in the simplified optical model potential 
$V(r)$ $=$ $-iW$ for $r$ $<$ $R$ (the ``black sphere 
model'') the zero-energy reaction cross section is an 
increasing function of $W$ for small $W$ only, and it 
decreases to zero for $W$ $\rightarrow$ $\infty$. 

In a previous work\cite{noi1} we have 
generalized the black sphere analysis showing that 
the inversion is associated with the formation 
of a sharp ``hole'' (i.e. a vacuum region with 
sharp boundaries, due to the annihilation) in 
the projectile wavefunction at small momenta. 
The underlying argument was not 
related with any specific model for the annihilation. 
It was stressed that this phenomenon was 
related to the transition from a semiclassical 
to a pure quantum, S-wave dominated, regime. 
We examined also more specific explainations for 
the inversion, however in the following  
we would like to further develop that general argument, 
relating it with the Heisenberg principle
$\delta k \delta x$ $>$ $1$ (in natural units). 

Both in Hydrogen and in heavier nucleus targets, 
the great bulk of the annihilations is supposed to take 
place within a region of thickness $\Delta$ 
$\approx$ 1 fm placed just out of the nuclear 
surface\cite{dov1,bonn1,brue91,aar,wei93}. 
Since the realized annihilation implies 
the statement ``$\bar{N}$ and nucleus at relative 
distance $r$ $\approx$ $R_{nucleus}$ 
defined within uncertainty $\Delta$'',
we expect strong deviations from semiclassical intuition 
at $k$ $<$ $1/\Delta$ $\sim$ 200 MeV/c.  

\section{The breaking of the saturation of the unitarity limit.}

We begin by spending a few words on the so-called ``black disk 
model'', that 
assumes complete flux remotion from the lowest partial 
waves and gives the unitarity limit for the total reaction 
probability. At very low energies, this model 
is a nonsense, for a well known\cite{ll1} 
limiting property of the phase shifts at zero energy.
Indeed, the black disk model 
assumes $\vert exp(2i\delta_o)\vert$ $=$ 0
for the S-wave phase shift $\delta_o$. 
But in the limit $k$ $\rightarrow$ 0 
any requirement of the kind  
$\vert exp(2i\delta_o)\vert$ $<$ $B$, where $B$ 
is a constant smaller than 1, means 
$Im(\delta_o)$ $\approx$ $-k Im(\alpha)$ 
$\rightarrow$ constant, i.e. 
$Im(\alpha)$ $\sim$ $-1/k$ $\rightarrow$ $-\infty$. 
This shows that the idea of complete flux remotion and 
the black disk model are ill-defined concepts at low energies. 
Of course, one can artificially put 
$\vert exp(2i\delta_o)\vert$ $=$ 0, 
but at small $k$ one will never be able to obtain this condition 
starting from a model with $confined$ interactions. 

So, in presence of a $very$ effective reaction mechanism, 
as annihilation is, we expect that a scale $k$ $\sim$ 
$k_b$ exists for the projectile momentum $k$ such that: 
(i) for $k$ $>>$ $k_b$ the reaction cross section assumes 
values which are close to the unitarity limit; 
(ii) for $k$ $<<$ $k_b$ we assist to a 
breaking of the saturation of the unitarity limit, i.e. 
the reaction cross section is much smaller that its 
unitarity limit value. 

Assuming that 
the main distortions in the entrance channel wavefunction 
are caused by the absorption, the uncertainty principle 
suggests $k_b$ $\sim$ $1/{\delta}r$, where ${\delta}r$ 
is the characteristic projectile path in nuclear matter.  
The consequent physics is very different depending whether 
this path is peculiar of a nucleus-projectile or 
of a nucleon-projectile underlying process. In this 
respect neutron induced nuclear reactions, and 
reactions like $\bar{N}$-nucleus annihilation or 
$K^-$-nucleus absorption, are the exact opposite. 
In the former case 
the underlying projectile-nucleon interactions 
are elastic, although their effect is destructive on the full 
nuclear structure. The reaction process contains the 
piece of information 
``the projectile and the nuclear center of mass are at 
relative distance $<$ $R_{nucleus}$, i.e. ${\delta}r$ 
$=$ $R_{nucleus}$''.  In the latter case the nucleon-projectile 
interaction is so inelastic that the path of the projectile 
in nuclear matter is $\Delta$ $\sim$ $R_{nucleon}$, and the 
reaction process contains the piece of information 
``relative distance $=$ $R_{nucleus}\pm\Delta$'', i.e. 
${\delta}r$ $\sim$ $\Delta$. 

In both cases the information implicitely contained in the 
fact that the reaction has happened is uncompatible with 
the statement ``the momentum of their relative motion 
was smaller than $1/{\delta}r$''. 
So, either the reaction 
can't happen or we must pay a price, in terms of 
large-momentum distortions 
of the projectile wavefunction. These distortions produce  
a large flux reflection, as we show below, that is the 
reason for the departure from the saturation of 
the unitarity limit. 

\section{The general mechanism.} 

We assume that 
the antinucleon-nucleus annihilation reaction is 
such a violent and effective process 
to make it necessary for the 
$\bar{N}$ wavefunction to be zero in all places 
where the value of the 
density of the nucleons is close to the nuclear matter value. 
In other words, as soon as the overlap between 
the distributions of probability 
for the antinucleon and for the target nucleons 
overcomes a certain threshold $<<$ 1 
the annihilation 
is supposed to take place, with the practical consequence 
that any consistent overlap of the projectile 
and target wavefunctions is forbidden. 
Most models\cite{dov1,bonn1,wei93} or phenomenological optical 
potential 
analyses\cite{brue91,aar,noi3} agree on this property. 

This produces a thin spherical shell of thickness 
$\Delta$ $\sim$ 1 fm (the exact size depends on 
the specific model) where the largest part of the 
annihilations is supposed to take place. We name 
it ``annihilation shell''. 
The internal surface of the annihilation shell roughly 
coincides with the surface of the target nucleus or proton, 
in agreement with 
the idea that $\bar{N}$ and nuclear matter densities 
can't overlap consistently. Depending on the model, 
the position of the 
external surface of the annihilation shell is related 
either with a minimum amount of overlap between 
antinucleon and nucleon densities required for 
annihilations, or with the range of a meson/baryon 
exchange between the annihilating particles. 
The target independence of $\Delta$, 
together with the Heisenberg principle, produces 
a target-independent annihilation cross section. To 
undestand how it realizes, we 
start with some easy 1-dimensional examples. 

We consider a $\bar{N}$ plane wave with momentum $\vec k$ 
$=$ $(0,0,-k)$ parallel to the $z$-axis.  
There is no interaction for $z$ $>$ 0, while for 
$z$ $<$ 0 absorption of the $\bar{N}$ flux is 
possible, according to some unknown mechanism.  
We don't know how it happens, but we know 
that most of the flux that enters the absorbtion 
region disappears within a range $\Delta$: 
$\vert \Psi(-\Delta)\vert$ $<<$ $\vert \Psi(0)\vert$.
The uncertainty principle implies that 
in this region the wavefunction has relevant 
components associated to a single particle 
momentum $k_z$ $\sim$ $1/\Delta$. 
A consequence of this is 
the obvious geometrical fact that for the 
absolute value of the logarithmic 
derivative of $\Psi$ we have 
$\vert \Psi'/\Psi\vert$ $\sim$ $1/\Delta$ in 
the damping range $-\Delta$ $<$ $z$ $<$ 0, and 
consequently also in $z$ $=$ $0-\epsilon$.  

For matching this value with the value of 
the logarithmic derivative on the 
positive $z$ side, we need both an incoming and a reflected 
wave. The general form of $\Psi(z)$ for $z$ $>$ 0 is 
$\Psi$ $=$ $\Psi_o sin[k(r-\alpha)]$ $\equiv$ 
$\Psi_{in} + \Psi_{out}$, with $\alpha$ 
complex to give account of the reactions. 
In general $\alpha$ is a function of $k$, 
however we can identify it with the $k$-independent scattering 
length since we are interested in the region of small $k$, 
and we assume that no resonances are present in the $k$-range 
that we consider. 

Below we report standard calculations, but it is easy 
to understand the relevant points in advance. 
For $z$ $>$ 0, $\vert\Psi_{in}\vert^2$ 
$\approx$ $\vert\Psi(z_p)\vert^2/4$, where $z_p$ is 
the lowest positive $z$ value where the periodical 
$\Psi$ attains an oscillation peak. 
$\vert\Psi(0)\vert^2$ $<<$ $\vert\Psi(z_p)\vert^2$ 
if $\vert \Psi'(0)/\Psi(0)\vert$ $>>$ $k$. 
As a consequence, for $\vert \Psi'(0)/\Psi(0)\vert$ $>>$ $k$ 
we have also 
$\vert\Psi(0)\vert^2$ $<<$ $\vert\Psi_{in}\vert^2$.  
In magnitude, 
$\vert\Psi(0)\vert^2/\vert\Psi_{in}\vert^2$   
$\sim$ $k^2\vert \Psi'/\Psi\vert^2$ 
$\sim$ $(k\Delta)^2$ at small $k$. 

The ratio between the value of $\vert \Psi(0)\vert^2$ 
and $\vert \Psi_{in}\vert^2$  
roughly coincides with the 
ratio between the absorbed and the incoming flux, 
or at least it represents an upper limit for this ratio. 
Indeed, only for $z$ $<$ 0 we may have flux 
absorption.

The ratio of the absorbed to the incoming flux will be 
a number of magnitude $\sim$ 1 only in the case where the 
condition $k$ $>>$ $1/\Delta$ is realized,  
because in this case the position $z_p$ 
will be close enough to the origin to have 
$\vert \Psi(0)\vert^2$ $\approx$ 
$\vert \Psi(z_p)\vert^2$. Then 
we are close to the saturation of the unitarity limit for 
the reaction: full flux absorption, possibly accompanied 
by elastically scattered diffractive flux (which originates 
in the interference between absorbed and incident waves). 
At $k$ $\sim$ $1/\Delta$ we start diparting from the 
saturation of the 
unitarity limit, and for $k$ $<<$ $1/\Delta$ we will 
be far from it. In the latter case the matching conditions 
associate a large $\vert\Psi'/\Psi\vert_0$ to  
a small flux absorption. As a by-product,  
elastic cross sections can be large, 
but they are refractive, not diffractive. 

If one wants to check the previous estimates with 
some calculations, one can normalize $\Psi$ for $z$ $>$ 0 
so to have 
$\Psi_{in}$ $=$ $e^{-ikz}$. Then $\Psi_o$ $=$ $e^{-2ik\alpha}$, 
and $\Psi_{out}$ $=$ $e^{ik(z-2\alpha)}$. Since the flux 
cannot be created, $Im(\alpha)$ $<$ 0. Then for $z$ $>$ 0 
\begin{equation}
\vert \Psi\vert^2\ =\ 1 + e^{4kIm(\alpha)} - 
2 e^{2kIm(\alpha)}cos\{2k[z-Re(\alpha)]\}. 
\end{equation}
In particular, when 
$k\vert Im(\alpha)\vert$ $<<$ 1 
$\vert\Psi\vert^2$ becomes $2-2cos\{2k[z-Re(\alpha)]\}$, 
so that also in presence of absorption 
$\vert \Psi(z_p)\vert^2/\vert\Psi_{in}\vert^2$ $\approx$ 4 
for $k$ small enough. When 
both $k\vert Im(\alpha)\vert$ $<<$ 1 
and $k\vert Re(\alpha)\vert$ $<<$ 1 are satisfied we have 
$\vert\Psi(0)\vert^2$ $\approx$ 
$(4k)^2 \vert\alpha\vert^2$ $<<$ 1, thus confirming that 
for $k$ small enough $\vert\Psi(0)\vert^2$ $\propto$ 
$(k\vert\alpha\vert)^2$.  
The logarithmic derivative of $\Psi$ in $z$ $=$ 0 is 
$k\cdot cotg(-k\alpha)$ $\approx$ $-1/\alpha$ at small $k$, 
so that ``$k$ small enough'' means $k$ $<<$ 
$\vert\Psi'(0)/\Psi(0)\vert$. 

The conclusions of the 
examined example may change if we 
consider a reaction region which is limited to 
$-z_o$ $<$ $z$ $<$ 0, i.e. for $z$ $<$ $-z_o$ no 
particle absorption is possible. 
We remark that $z_o$ represents 
the size of the region where reactions are possible, 
while $\Delta$ is the range needed for the projectile 
wavefunction to pass from $\Psi$ $\approx$ 
$\Psi_o sin[-k(z-\alpha)]$ to $\Psi$ $\approx$ 0. 
We must distinguish 
the two cases where $z_o$ is smaller or larger 
than $\Delta$. 
In the former case the absorption is 
proportional to the thickness $z_o$ of 
the reaction region. But a saturation condition is reached 
when $z_o$ becomes larger than $\Delta$, 
and for any $z_o$ $>>$ $\Delta$ the 
conclusions will be 
the same as in the case $z_o$ $=$ $\infty$. 
It is now useful to notice that for $z_o$ 
$>>$ $\Delta$ nothing would be changed by the 
introduction of the additional boundary 
condition $\Psi(-z_o)$ $=$ 0. This constraint 
obliges one to take into account 
the reflected wavefunction inside the reaction region, 
i.e. that component of $\Psi$ whose absolute value 
increases at increasing negative $z$ inside the 
reaction region. But for $z_o$ $>>$ $\Delta$ this component 
is very small and can be neglected. 

The latter situation with the additional ``reflection''
condition  in $-z_o$ corresponds to the 1-dimensional 
reduction of the 3-dimensional 
problem of $\bar{N}N$ and $\bar{N}-$nucleus annihilation, 
because the damping of the projectile wavefunction 
takes place on a space scale which is short enough 
to prevent antinucleons from reaching the origin 
with any target. From a mathematical point of view 
the situation is identical in the two cases, after 
substituting $\Psi(z+z_o)$ with $r\Psi(r)$. 

In treating the problem, 
initially we neglect the role of a real $strong$ attracting 
potential. The modifications that it introduces will 
be considered in a further section. 
We define $R_m$ and $\Delta$ such that 
practically all of the annihilations are supposed to 
take place at $r$ values comprised between $r$ $=$ 
$R_m$ and $r$ $=$ $R_m-\Delta$. We assume $R_m$ 
as a reasonable matching radius,  
satisfying the two conditions: (i) 
for $r$ $>$ $R_m$ the 
oscillations of $\chi(r)$ $\equiv$ $r\Psi(r)$ are 
mainly controlled by the sum of 
kinetic and Coulomb potential energy, and 
the distortions of $\chi$ due to the absorption 
are negligible; (ii) at smaller radii 
the situation becomes the opposite within a range 
$<<$ $1/k$. The interactions  
$directly$ responsible for the annihilation have 
range $R$ and decay exponentially for $r$ $>$ $R$
according to some $exp(-r/r_o)$ law    
(e.g., for a Woods-Saxon potential $R$ is the radius and 
$r_o$ the diffuseness). 
Depending on the model, $R_m$ 
is normally 0.5-1 fm larger than $R$, suggesting that 
the relevant processes take place in the exponential 
tail of the annihilating forces. 
Clearly $1/k$ defines the ``scale of space resolution'' 
in the problem, and the following considerations can 
be applied for $k$ $<<$ $1/r_o$ only. 

Summarizing, in our problem we assume both the 
range $r_o$ characterizing the exponential damping of 
the inelastic interaction and the 
thickness $\Delta$ of the annihilation shell 
to be much smaller than 
$1/k$, and assume the reasonable matching radius $R_m$  
to be larger than $\Delta$. 
With the previous definitions and assumptions, 
all the things that we have written about the 
``$z-$problem with reflection condition'' can 
be repeated word by word after substituting $z+z_o$ with $r$,  
$\Psi(z+z_o)$ with $\chi(r)$ $\equiv$ 
$r\Psi(r)$, while $r$ $=$ $R_m$ 
corresponds to $z$ $=$ 0 and $r$ $=$ 0 to 
$z$ $=$ $-z_o$. 
More properly however, $k$ is the 
wavenumber produced at $r$ $=$ $R_m$ by both the 
kinetic and the Coulomb potential energy. 

The saturation condition 
is expressed by $R_m$ $>$ $\Delta$ $\approx$ 1 fm, 
and seems to be realized, as above discussed, 
in antinucleon annihilation on all possible targets, from 
proton to heavy nuclei. It implies that the 
reflected flux is negligible $inside$ the proton/nucleus 
target. The uncertainty principle assures that 
the dominating momentum components 
inside the reaction  range are $\sim$ $1/\Delta$.  
When this is transferred to the $\bar{N}$ 
wave it means $\vert \chi'/\chi\vert_{R_m}$ $\sim$ 
$1/\Delta$, with large flux reflection for 
$k\Delta$ $<<$ 1. 

In the S-wave 1-dimensional reduction of 
the 3-dimensional scattering problem the reflected 
wave is a composition of both the scattered 
and of the untouched initial wave. 
The disappeared flux corresponds to 
inelastic reactions, and the ratio of this flux to 
the incoming one is $\sim$ $(k\Delta)^2$ for 
$k\Delta$ $<<$ 1, in agreement with the previous 
example. A part of the reflected flux will 
correspond to elastic reactions, which are not diffractive 
because we are very far from the unitarity limit.  
The fact that the 
above ratio of the absorbed to the incoming 
flux tends to zero for $k$ $\rightarrow$ 0 is not in 
contraddiction with a finite reaction rate, 
but target details are lost once $k$ $<$ $1/\Delta$. 

\section{Predictions.} 

It is easy to estimate upper limits for 
the complex scattering length $\alpha$ 
with the condition 
\begin{equation}
\vert \chi'/\chi\vert_{R_m-\epsilon}\  
\approx\  1/\Delta.
\end{equation}
Using $\vert \chi'/\chi\vert_{R_m+\epsilon}$ 
$=$ $k\cdot cotg[k(R_m-\alpha)]$ one finds, in the limit 
$k$ $\rightarrow$ 0, 
\begin{equation}
\vert \Delta\vert^2\ \approx\  
[R_m-Re(\alpha)]^2+[Im(\alpha)]^2,
\end{equation}
that implies: 

$\vert Im(\alpha)\vert $ $\approx$ $\Delta$ 
or smaller, 

$Re(\alpha)$ is positive and 
comprised in the range $R_m\pm\Delta$. 

The consequence of this are: 

1) $\Delta$ (rather than $R_{nucleus}$) is the relevant 
parameter for the low energy reaction probability, which is 
proportional to $\vert Im(\alpha)\vert$. For 
$k$ $<<$ 100 MeV/c and 
for $\bar{n}$ projectiles the reaction probability 
should be roughly the same for any target nucleus radius, 
as far as the reaction is S-wave dominated (so, for 
$k$ $<<$ $100/A^{1/3}$ MeV/c), with magnitude 
$\pi\Delta/k_{cm}$ $\approx$ 
$6000 \Delta/k_{cm}$ mb (with $\Delta$ in fm and $k$ 
in MeV/c).  
For $\bar{p}$ projectiles the differences will be 
mostly due to the Coulomb effects, which have 
been estimated elsewhere\cite{noi2,cp1}. Both with 
$\bar{n}$ and with $\bar{p}$, 
$Im(\alpha)$ $\sim$ 1 fm (or smaller) for all nuclear 
targets. 

2) $Re(\alpha)$ $\sim$ $+R_m$ $\sim$ $+R_{nucleus}$
means an $\bar{n}-$nucleus total elastic 
cross section $\sim$ $4\pi R_{nucleus}^2$, 
and its positive sign is characteristic of a $repulsive$ 
interaction. Accordingly, 
the zero energy $\rho-$parameter $=$ 
$Re(\alpha)/Im(\alpha)$ is negative.   
We can estimate $Re(\alpha)$ $\sim$ 1 fm and $\rho$ 
$\approx$ $-1$ for light nuclei,  
$Re(\alpha)$ $\sim$ 1.3 $A^{1/3}$ fm and $\rho$ 
$\approx$ $-A^{1/3}$ for heavy nuclei. 
Again, Coulomb 
effects enhance the total elastic cross section 
in the $\bar{p}$ case\cite{noi3}. 

3) If one can identify subsets of $\bar{N}N$ annihilation 
events 
which are supposed to be characterized by different 
$\Delta-$parameters, the consequent low-energy cross sections 
should scale accordingly. E.g., $\bar{p}p$ $\rightarrow$ 
$2 \pi$ and $\bar{p}p$ $\rightarrow$ $2 K$ have been 
demonstrated to be characterized by different space scales, 
because of the different mass of the final states\cite{pik}. 
If the characteristic 
annihilation distances, measured at $k$ $>>$ 200 MeV/c or  
estimated by some model,  
are $\Delta_1$ and $\Delta_2$, 
the ratio between the corresponding annihilation rates 
should be of magnitude $(\Delta_1/\Delta_2)^2$ at 
very small momenta. 

For all those reactions (e.g. $K^-$ absorption 
on nuclear targets), where the absorption range inside 
nuclear matter is $\sim$ 1 fm, the same considerations 
apply. Relevant deviations from the previous predictions 
should be attributed to peculiarities of the external 
tail of the nuclear density (e.g. a longer  
tail in deuteron or $^3$He, or a different 
proton/neutron composition at the surface). 
In the special case of neutron-halo nuclei the 
presence of a very long range tail in the nuclear matter 
distribution removes the basic assumptions of this work. 

\section{The role of elastic attracting potentials} 

In presence of a real attracting potential surrounding 
the annihilation shell the actual 
zero-energy momentum at $R_m$ is determined by the 
potential energy. We must consider two very different 
cases, i.e. strong or Coulomb interactions. 

A strong elastic 
potential has nuclear characteristic range, so it does 
not escape the previous general considerations. 
Now the external surface of the annihilation shell 
should be displaced to include the region where the 
distortions of the projectile wavefunction  
of elastic origin are relevant.  
This may increase $\Delta$ up to 
2 fm\cite{noi3,dov1,aar,brue86}. However, 
the convergence of the $\bar{p}p$, $\bar{p}D$, $\bar{p}^4$He 
and $\bar{p}^20$Ne 
annihilation cross sections to similar values at small momenta,  
all corresponding to scattering lengths $<$ 1 fm (after 
subtracting Coulomb effects) suggests that $\Delta$ is smaller 
than 1 fm. 

The Coulomb potential has atomic range and so 
escapes all the previous considerations.  
In $\bar{p}p$ annihilations, Coulomb 
forces fix a minimum $\bar{p}$ kinetic energy of magnitude 
1 MeV at the proton surface, corresponding to a 
momentum 40 MeV/c, that represents 
a scale for the true zero 
energy momentum we have to consider. At much smaller 
momenta all the modifications that we observe 
are due to electromagnetic or atomic effects. 
With nuclear targets, 
the Coulomb energy at the nuclear surface increases 
proportionally to $Z/Z^{1/3}$ $=$ $Z^{2/3}$, so 
the corresponding zero-energy momentum increases 
proportionally to $Z^{1/3}$. With very heavy nuclei the 
Coulomb momentum starts becoming comparable 
in magnitude to the Fermi momentum, introducing a completely 
different physics. Apart from this, Coulomb forces 
produce a large enhancement of the  
reaction and elastic cross sections by focusing the 
$\bar{p}$ wavefunction on the nucleus. This effect 
is widely discussed in other 
works\cite{noi3,noi2,cp1}. 

\section{Conclusions}

We have shown that, within those models where the 
annihilation 
probability is large enough to prevent a consistent overlap 
between the projectile and the target wavefunctions, the 
antinucleon-nucleus annihilation cross section is largely 
target-independent, apart for Coulomb effects. 
The cause of this behavior is the 
quantum uncertainty principle, together with the fact that 
on most of the nuclear targets the process is characterized 
by the same value of the parameter $\Delta$ $\sim$ 1 fm.  
$\Delta$ is the thickness of the spherical shell 
surrounding the nucleus where the bulk 
of the annihilations are supposed to take place. 
For the scattering length $\alpha$ we have estimated 
$Im(\alpha)$ $\approx$ $\Delta$, while $Re(\alpha)$ 
is positive and 
roughly coincides with the larger between the nuclear 
radius and $\Delta$.  
We have also suggested that the ratio between the low 
energy annihilation rates relative to selected final states with 
different characteristical annihilation distances $\Delta_1$ 
and $\Delta_2$ should be $\Delta_1/\Delta_2$.

\end{document}